**Sizable electromagnetic forces in parallel-plate metallic cavity**

S. B. Wang, [1,*] Jack Ng, [1] H. Liu, [2] H. H. Zheng, [1] Z. H. Hang, [1] and C. T. Chan[1,†]

[1]*Department of Physics and William Mong Institute of Nano Science and Technology, The Hong Kong University of Science and Technology, Clear Water Bay, Hong Kong, China*

[2]*Department of Physics, National Laboratory of Solid State Microstructures, Nanjing University, Nanjing 210093, People's Republic of China*

**ABSTRACT**

Using a boundary element method to calculate the electromagnetic fields and the Maxwell stress tensor method to compute the electromagnetic forces, we investigate electromagnetic wave induced forces acting on a pair of identical metal plates that forms an electromagnetic resonance cavity. Different frequency regimes are considered, from infrared frequencies with micron scale structures down to the microwave regime which involves millimeter scale structures. We found that at both length scales, the electromagnetic wave induced forces can be significantly stronger than the usual photon pressure exerted by a laser beam if the cavity is excited at resonance although the mechanisms that underlie the strong force are different at different length scales. In the infrared frequency regime, the strong force is induced by field penetration into the metal, whereas in the microwave regime, the electromagnetic force is induced by the leakage of electric field at the edges. At both frequency scales, we compare the results we obtained for Au metal plates with fictitious perfect electric conductor plates so as to understand the effect of field penetration. We also showed that a transmission line model can give simple expressions that can capture the essence of the physics. The effect of surface corrugation and surface roughness is also investigated, and we find that corrugation/roughness generally induces attraction between the plates.

# I. INTRODUCTION

The invention of the laser paved the way for particle manipulation by light induced forces.[1,2,3,4] Nowadays, optical force is used extensively to manipulate small particles,[5,6,7,8,9,10,11,12,13,14] and it is intensively investigated in special systems such as waveguides[15,16,17,18,19,20,21,22] and nanowires.[23,24] The term optical force typically refers to the forces induced by visible or near infrared (IR) radiation, which accounts for only a small part of the electromagnetic spectrum. We know that static fields can also manipulate particles, a good example being electrophoresis. We note that the optical force can be enhanced by the excitations of surface plasmons at visible or IR frequencies, but this important phenomena is absent at microwave frequencies. Is it possible to achieve reasonably large forces at intermediate frequencies such as microwave frequencies? The purpose of this paper is to show that electromagnetic forces and pressure induced by the resonance in a metal sandwich configuration can be sizable in both high (near IR) and low (microwave) frequency regimes using the same geometric configuration and the same material.

We shall explore the electromagnetic forces induced by external illumination for a system consisting of two parallel metallic plates separated by an air gap and the basic configuration of our system is depicted in Fig. 1. The plates are under the illumination of an electromagnetic plane wave whose wave vector is normal to the surface of the plates and the polarization is such that the magnetic field points in the *z*-direction, as shown in Fig. 1(b). In the microwave regime, such a metal plate sandwich configuration is similar in structure to the "mushroom" elements of the high-impedance ground planes,[25] except that our system has no metal vias connecting the metal plates. In the plasmonic regime, such metal sandwich structure has been used to facilitate the realization of negative effective permittivity and negative effective permeability at high frequencies.[26,27]

We note under the illumination of electromagnetic wave, such a parallel-plate system sustains two kinds of coupling modes: the symmetric mode and the anti-symmetric mode.[24] The former mode is characterized by a pair of parallel currents induced on the two plates while the latter is characterized by a pair of anti-parallel currents. For the symmetric mode, the fields of the two plates tend to cancel each other and most of the electromagnetic energy stays outside the cavity. The optical force corresponding to the symmetric mode is much smaller than that of the anti-symmetric mode in our cases (see Appendix). In the following discussion, we focus on the anti-symmetric mode which gives forces that are two to three orders of magnitude stronger than that of the symmetric mode.

By Faraday law, the time-varying magnetic flux between the plates generates an electromotive force which drives the electrons to move synchronously, leading to currents depicted pictorially by the yellow arrows in Fig. 1(b). At certain frequencies, the periodic oscillation of the induced currents will be at resonance with the external driving fields and very strong fields will be induced between the plates. At resonance, the anti-parallel currents induced on the two plates will exert a repulsive magnetic force on the plates. At the same time, current oscillations lead to charge accumulation in the plates as illustrated schematically in Fig. 1(b), and this generates an electric Coulomb attraction. The net or total optical force due to the resonance is the sum of the repulsive magnetic force and the attractive electric force. In other words, the attractive/repulsive nature of the force is determined by the competition between the electric field effect and the magnetic field effect. In a previous paper,[28] we reported that a strong attractive force between a pair of plasmonic plates can be induced by the kinetic energy of the electrons in IR range. In this article, we consider and compare the optical forces in both the high frequency (IR) and low frequency (microwave) regimes and we will employ Au as the prototype metal. We will also give the corresponding results if the metal plates were replaced by fictitious perfect electric conductors (PECs). We find that if there is no field penetration into the metal and if we can ignore the fringe effect, the electric field induced attraction and the magnetic field induced repulsion essentially cancel each other, leaving behind a small residual force due to light reflection on the front surface. In that ideal limit, the strong resonant fields do not generate a strong force as the magnetic field effect and the electric field effect oppose each other even though the magnitude of fields can be very strong due to resonance. When fringe effects are taken into account, the field leakage at the edges affects the electric field more than the magnetic field. This is because the electric field is stronger near the edges while the magnetic field is stronger near the center of the sandwich. And the leakage effect leads to a smaller attractive force induced by electric field. But if the field penetrates the metal, a certain portion of the magnetic energy is stored inside the metal as the kinetic energy of the electrons, and the total Faraday magnetic energy becomes small, so that the magnetic field induced repulsion is diminished. The result is that fringe effect decreases attraction and field penetration suppresses repulsion. In the microwave frequency regime, the field hardly penetrates the metal and we will see that the resonant electromagnetic force is mainly dominated by the fringe effect and hence the force is repulsive while resonant optical force becomes attractive at high (e.g. IR) frequencies as the field starts penetrating the metal.

At each frequency scale, we will compare the results we obtained for Au metal plates with those for ideal PEC plates so as to understand the effect of field penetration. And for a better heuristic understanding of the physics, we employed a transmission line model which gives simple expressions that can capture the essence of the physics.

We will see that rather strong electromagnetic forces, which can be hundreds of times stronger than the usual photon pressure due to reflectance on a metal surface, can be induced by the electromagnetic resonances in both frequency regimes. We note in particular that the microwave induced force is not small so that it is possible to use conventional microwave sources to obtain measurable forces with a relatively low output power. At high (IR or optical) frequencies, one can achieve strong optical forces either by generating strong fields via the excitation of some sort of resonances or by focusing light to a small spot. Manipulating materials using strongly focused light at optical frequencies is routinely practiced in optical tweezers.[29,30,31] However, it is much more difficult to focus the microwave energy into a small volume. Even when one focuses the microwave energy to a diffraction limited spot, the resulting intensity is still typically orders of magnitude weaker than its optical analogue, and it is not sufficient for manipulation. For this reason, in addition to focusing, inducing electromagnetic forces through resonance effects will be necessary in the microwave regime. We do want to emphasize again that the physics here is more subtle than mere field enhancement. While resonances generally induce strong forces [32,33] because of field enhancement, the electromagnetic forces have opposite signs for electric and magnetic fields in the metal sandwich configuration and the field enhancement *per se* cannot account for the strong force in this system. Additional mechanism (either field penetration or fringe effect) is needed to suppress the effect of one of the fields (e.g. field penetration selectively suppresses the magnetic field effect), leaving behind the effect of the other field to realize a strong force.

In what follows, we will first introduce in the Methodology section our computational methods and an analytical model which satisfactorily explains the simulation results. In the Results and Discussion section, we will show results for the optical force/pressure at the micron scale, comparing the results obtained for Au with those for PEC plates. We then consider the electromagnetic force/pressure at the millimeter scale, with resonances in the microwave frequency regime. The effect of surface corrugation and roughness is then considered.

## II. METHODOLOGY

### A. Geometry and setting

We consider a pair of identical metallic plates separated by a distance $d$, each characterized by thickness $t$, length $l$, and width $w$ (length along the $z$-direction), they form a two-dimensional system as shown in Fig. 1(a). In this paper, the metallic plates are assumed to be made of gold, and we will also present corresponding results for PEC plates to help understand the physics. We will consider two length scales. In part III.A, we consider the wavelength and the size of the plates to be on the order of microns, whereas in part III.B, the scale increases to millimeter, corresponding to resonances at microwave frequencies. Unless otherwise noted, the incident electromagnetic wave has the form of a plane wave

$$\mathbf{E}_{in} = E_0 \hat{y} e^{ikx}, \quad (1)$$

where the $k$-vector of the incident wave is perpendicular to the flat side of the plates, as shown in Fig. 1(b). The polarization of the incident field is along the $y$-axis, with the coordinate system defined in Fig. 1(a).

**B. Numerical computation: boundary element method and Maxwell stress tensor**

In the following (Figures 2 to 7), we will show numerically computed optical pressures acting on our system. There are two metal plates in our system, and we will call the plate on the left (see Fig. 1(b)) the front plate, on which the incident light coming from the left impinges, and we call the plate on the right the back plate. The numerically calculated optical pressure we will show is the time averaged electromagnetic force acting on the back plate divided by the area of the plate. We would be primarily interested in the optical force/pressure acting on the plates when the system is in resonance and in that case, the optical pressures acting on the front plate and back plate are nearly equal, with the difference being the photon pressure due to reflection on the front metal plate which is small (typically less than 1% of the total optical pressure acting on one plate). The numerical computation of the optical forces involves two steps. First, the boundary element method (BEM)[34,35,36] is applied to solve the Maxwell equations with the standard boundary conditions, which gives the electromagnetic fields. The detail of the BEM we employed is described in Ref. 34. With the fields given by BEM, the total time averaged optical force acting on the back plate is then calculated by using the standard Maxwell stress tensor[37] approach, and the area averaged optical pressure is obtained by dividing the force by the specific area of the plate. We shall refer to the area averaged optical pressure simply as optical pressure hereafter.

**C. Analytical distributed transmission line theory**

While the numerical solution of the Maxwell equations can give us numerical results for a specific configuration as precise as we wish, it would be useful if we can use a simple model to get an intuitive understanding of the underlying physics. To that end, we employ a transmission line[38,39] model that can explain the numerical results qualitatively and give some insight to the physics. The pair of parallel plates can be treated as an open-end transmission line, which consists of RLC circuit units, as shown in Fig.1(c). In this model, the non-uniform distribution of charges and currents on the plates are accounted for by the distributed effect of the transmission line. For an incident electromagnetic plane wave (Eq. (1)), the position-dependent voltage $V(y)$ and the position-dependent current $I(y)$ are governed by the telegraph equations[40,41]

$$\frac{dV(y)}{dy} + (R - i\omega L)I(y) = i\omega\mu_0 \int_{x_1}^{x_2} \hat{z} \cdot \mathbf{H}_{in}(x)dx, \tag{2}$$

$$\frac{dI(y)}{dy} + (G - i\omega C)V(y) = i\omega(G + C)\int_{x_1}^{x_2} \hat{x} \cdot \mathbf{E}_{in}(x)dx. \tag{3}$$

Here $\mathbf{H}_{in}(x)$ is the incident magnetic field and $\mathbf{E}_{in}(x)$ is the incident electric field; $L$ and $C$ are the per-unit-length inductance and capacitance for the parallel-plate system, $R$ is the resistance per unit length and $G$ is the conductance per unit length of the air gap between the metal plates, and $d_{eff} = x_2 - x_1$ is the effective separation between the plates (see Appendix). In Eq. (3), $G$ and $\hat{x} \cdot \mathbf{E}_{in}$ are zero for the particular configuration under consideration. Differentiating Eq. (3) with respect to $y$, and then substituting it into (2), we obtain the differential equation for $I(y)$, which can be solved by applying the boundary condition $I(0) = I(l) = 0$, i.e. the currents at the ends of the plates are zero. The solution is given by

$$I(y) = \frac{B}{A^2}\left[1 - \cos(Ay) - \tan(Al/2)\sin(Ay)\right], \tag{4}$$

where $A = \sqrt{\omega^2 CL + i\omega RC}$ and $B = -\mu_0 \omega^2 C \int_{x_1}^{x_2} \hat{z} \cdot \mathbf{H}_{in}(x)dx$.

When $R$ is small, the resonance condition, characterized by having a finite output at vanishing input, can be inferred from (4) as:

$$Al = (2n+1)\pi, \tag{5}$$

where $n$=0, 1, 2, …, and the resonant frequency is given by

$$\omega_n = \frac{2\pi c}{\lambda} = \frac{(2n+1)\pi}{\sqrt{LCl^2}}. \tag{6}$$

We are only interested in the lowest order resonance, namely $n$=0 in (6). Note that higher order resonances also exist in our system, but their corresponding electromagnetic forces are weaker than that of the $n=0$ resonance (see Appendix). At the $n=0$ resonance, the amplitude of the current on each plate is approximately sinusoidal (see Appendix):

$$|I(y)| \approx \left| \frac{B}{A^2} \tan(Al/2) \sin(Ay) \right|. \tag{7}$$

We note that Eq. (6) is exact only for the ideal transmission line or when $d/l \to 0$. For real transmission line with a non-zero $d/l$, the electric field will leak at the edges of the plates as shown in Fig. 1(b), and the field leakage will result in an additional capacitance. In order to account for this additional capacitance arising from the fringe effect, we designate an effective length to the capacitance,[42] i.e. replacing $Cl$ with $Cl_{eff}$, where $l_{eff} = l(1 + \frac{\alpha d}{l} + ...)$ contains the first order correction of the capacitance in $d/l$, and $\alpha$ is the coefficient of the first order term. With that the resonant frequency (6) generalizes to, up to first order in $d/l$:

$$\omega_n = \frac{2\pi c}{\lambda} = \frac{(2n+1)\pi}{l\sqrt{LC(1+\alpha d/l)}}. \tag{8}$$

In our analytical calculation, we are only interested in the $n=0$ mode, and the constant $\alpha$ will serve as a fitting parameter. While $\alpha$ does not significantly alter the resonant frequency (because the zero-th order term dominates), we shall see that for PEC, the optical force is derived from the fringe effect and thus depends on $\alpha$.

If the electromagnetic energy inside the cavity is mainly contributed by the resonant mode, the optical force induced on the plates by the incident electromagnetic radiation is given by force formula:[16, 33]

$$F = -\frac{\partial U}{\partial d} \approx -\frac{\partial U}{\partial \omega_0} \frac{\partial \omega_0}{\partial d} \approx -\frac{U}{\omega_0} \frac{\partial \omega_0}{\partial d}, \tag{9}$$

where $\omega_0$ is the resonant frequency corresponding to the $n=0$ mode and $U$ is the total energy stored in the system. We note that Eq. (9) only accounts for the "inter-plate force" between the plates derived from the resonance excitation of the cavity. This expression does not include the photon momentum transfer due to photons hitting the front plate. As the resonance force is typically hundreds of times stronger than the simple momentum transfer, the latter can be ignored in the discussion of the physics, although in the numerical calculations, "everything" is included within the Maxwell stress tensor formulation. At resonance, the electric energy and the magnetic energy are equal, and thus $U = 2 \times \frac{L}{2} \int_0^l I_{eff}^2(y) dy$, where the time-averaged effective current is given by

$$I_{eff}(y) = \frac{1}{\sqrt{2}} \sqrt{I(y) \cdot I(y)^*}, \tag{10}$$

with the asterisk denotes the complex conjugate and $L$ is the total inductance per unit length. We shall treat the case of Au and ideal PEC plates separately.

*1. Au plates in the plasmonic regime (at the micron scale)*

Let us consider the plates to be made of Au, and we consider the plasmonic regime in which we can model Au reasonably well in the IR regime using the Drude model[37]

$$\varepsilon_r = 1 - \frac{\omega_p^2}{\omega^2 + i\omega_t \omega}, \tag{11}$$

where[43] $\omega_p = 1.37 \times 10^{16}$ rad/s and $\omega_t = 4.084 \times 10^{13}$ rad/s. The relevant length scale would be the micron. The Maxwell's equation for the metal is

$$\nabla \times \mathbf{H} = -i\omega(\varepsilon_0 + i\frac{\sigma}{\omega})\mathbf{E} = -i\omega\varepsilon_r\varepsilon_0\mathbf{E}. \tag{12}$$

From (11) and (12), we obtain the complex resistivity as

$$\rho = \frac{1}{\sigma} = \frac{\omega_t - i\omega}{\varepsilon_0 \omega_p^2}. \tag{13}$$

The per-unit-length impedance of a single plate is given by

$$Z = \frac{\rho}{2\delta w} = \frac{\omega_t}{2\delta w \varepsilon_0 \omega_p^2} - \frac{i\omega}{2\delta w \varepsilon_0 \omega_p^2} = \bar{R} - i\omega\bar{L}_e, \tag{14}$$

where $\delta$ is the penetration depth or skin depth, $\bar{R}$ is the per-unit-length resistance for one plate and $\bar{L}_e$ is the per-unit-length kinetic inductance for one plate.[44,45,46] Under the condition $\omega \gg \omega_t$ the penetration depth is given by $\delta = c/\omega_p$,[46] where $c$ is the speed of light in vacuum. Since there are two plates in the system, the per-unit-length resistance and kinetic inductance are given by $R = 2\bar{R} = \omega_t / (\delta w \varepsilon_0 \omega_p^2)$, $L_e = 2\bar{L}_e = 1/(\delta w \varepsilon_0 \omega_p^2)$. The total per-unit-length inductance (sum of Faraday and kinetic inductance) is $L = L_m + L_e = \mu_0 d/w + 1/(\delta w \varepsilon_0 \omega_p^2)$, where the first term is the conventional parallel-plate magnetic inductance per unit length. The per-unit-length capacitance of the parallel-plate system is $C = \varepsilon_0 w/d$. With $R$, $L$ and $C$ defined, we apply (9) to obtain the electromagnetic force

$$F = \frac{U}{2} \frac{\alpha \mu_0 d^2 - L_e wl}{(L_e w + \mu_0 d)l^2[d/l + \alpha(d/l)^2]} \approx \frac{U}{2d(L_e + L_m)} \left( -L_e + \frac{\alpha \mu_0 d^2}{wl} \right). \quad (15)$$

In deriving Eq. (15) (see Appendix), we retained only the zero-th order and first order terms in $d/l$, consistent with our Taylor expansion in (8). Eq. (15) indicates that while the leading zero-th order term is attractive (negative), the first order term is repulsive. We note that the first order term is proportional to $\alpha$ and it corresponds to the fringe effect. We also note that the magnitude of the force is proportional to the total energy stored in the system ($U$).

*2. PEC plates*

We now consider the ideal case in which the plates are made of PECs for comparison. PECs are good approximations in the microwave frequency regime, corresponding to the scale length of millimeters. Considering the ideal PEC limit also gives us deeper insight into the physics in the micron length scale. In the case of PEC plates, the absence of the Ohmic loss implies that the only loss mechanism is the radiation loss. It is difficult to obtain an accurate value for the radiation loss without using a full wave numerical simulation tool. Here, we estimate the radiation loss by using the approach in Ref. 47.

We outline the approach in the following. The total radiation resistance is defined through $R_r^{tot} I_{eff}^2 = W$, where $W$ is the time averaged radiation power, which can be calculated by

$$W = \frac{1}{2} \int_\Omega \text{Re}(\mathbf{E} \times \mathbf{H}^*) \cdot d\mathbf{S}, \quad (16)$$

where the integration domain is over a surface $\Omega$ that lies in the radiation zone (i.e. infinity) and encloses the whole source region. In Eq.(16), the electromagnetic fields can be computed from the vector potential

$$\mathbf{A}(\mathbf{r}) = \mu_0 \int_{V'} \frac{\mathbf{J}(\mathbf{r'})e^{ik\bar{r}}}{4\pi\bar{r}} dV' \approx \mu_0 \frac{e^{ikr}}{4\pi r} \int_{V'} \mathbf{J}(\mathbf{r'}) e^{-ikr'\cos\psi} dV' \qquad (17)$$

under the Lorentz gauge

$$\nabla \cdot \mathbf{A} = -\varepsilon_0 \mu_0 \frac{\partial \phi}{\partial t}. \qquad (18)$$

Here $\mathbf{J}(\mathbf{r'})$ is the current density at position $\mathbf{r'}$ and $\bar{r} = |\mathbf{r} - \mathbf{r'}|$. We made use of the fact that $\bar{r} = \sqrt{r^2 + r'^2 - 2rr'\cos\psi} \approx r - r'\cos\psi$, where $\psi$ is the angle between $\mathbf{r}$ and $\mathbf{r'}$. We assume a sinusoidal surface current $\mathbf{J} = \hat{y} J_0 \sin(\pi y/l)[\delta(x-x_1) - \delta(x-x_2)]$, as expected from Eq. (7) when the system is at resonance. The integral (17) can then be evaluated numerically. Finally, we get the effective per-unit-length radiation resistance as $R_r = W/(lI_{eff}^2)$ (as it is difficult to determine the exact per-unit-length radiation resistance, we take the average for simplicity).

As there is no field penetration, the inductance (per unit length) of the PEC system is the Faraday inductance $L = L_m = \mu_0 d/w$ and its capacitance (per unit length) is $C = \varepsilon_0 w/d$. With $R_r$, $L$ and $C$ known we can calculate the current distribution through Eq. (4) and the total energy $U$. Then we can obtain the expression of optical force in the PEC case with Eq. (8) and (9) as

$$F = \frac{\alpha U}{2l}, \qquad (19)$$

which indicates that without fringe effect ($\alpha = 0$) there will be no optical force due to the resonance in the system if the plate is made of PEC. We shall see later from numerical results that $F \propto l^2$ for plane wave incidence and Eq. (19) implies that $U \propto l^3$.

### III. RESULTS AND DISCUSSIONS
#### A. Electromagnetic forces at the micron scale: Au vs PEC

According to Eq. (15), the first term, which is proportional to kinetic inductance $L_e$, gives a negative (attractive) force and the second term, which comes from the fringe effect, gives a positive (repulsive) force contribution. If the plates are made of PEC where $L_e=0$, the first term would disappear so that the electromagnetic force should be repulsive, and the repulsion is derived from the fringe effect as parameterized by the constant $\alpha$. We note that $\alpha$ is generally positive since the fringe effect is basically a field leakage effect. In the plasmonic regime for the Au plate, the field can penetrate the metal, and we must take the kinetic inductance $L_e$ into account, and the force can become negative (attractive) if the inductance term overwhelms the fringe effect. This is indeed the case as shown in Fig. 2, in which we show the optical pressures at the micron length scale calculated numerically with the BEM and the Maxwell stress tensor method. The system parameters are specified in the figure captions. In Fig. 2(a) we plot the calculated optical pressures for the PEC case and Fig. 2(b) we show the pressures for the Au case. We note here again that the optical pressure corresponds to the optical force per unit area acting on the back plate. At this length scale, the field should penetrate the metal and hence the PEC results in Fig. 2(a) are just for comparison with those shown in Fig. 2(b) so that we can understand the consequence of field penetration in the Au plate. It is clear from Fig. 2(a) that for the PEC system the peak value of the pressure, defined as

$$P = \frac{F}{lw}, \tag{20}$$

where $F$ is defined in Eq. (19), increases linearly with respect to the length of the plates ($P \propto \lambda \propto l$). The optical force is positive, as predicted if kinetic inductance is zero. However, the numerical results in Fig. 2(b) show that once we consider real Au as parameterized by the Drude model, the field penetrates the metal and hence there is a kinetic inductance, and the kinetic inductance dominates the fringe effect and the force changes sign to become negative (attractive). We note that for a single electromagnetic plane wave illuminating a single metal plate, the magnitude of electromagnetic photon pressure is less than 6.7 Pa/(mW/μm$^2$) while the optical pressure here (Fig. 2(b)) can reach ~ 10$^3$ Pa/(mW/μm$^2$) at resonance. Consequently, the resonance created by a pair of plasmonic plates enhanced the optical pressure by 2 to 3 orders of magnitude, as shown in Fig. 2(b). Instead of a monotonic increase of optical pressure as in the case of PEC, there is an optimal $l$ for "plasmonic" Au plates, in which the optical pressure is the largest. This optimal $l$ appears to be around 0.4μm and with a magnitude of approximately $-2.3\times10^3 \text{Pa/(mW/μm}^2)$. The existence of an optimal value of $l$, as told by Eq. (15), is due to the competition of the kinetic inductance and the fringe effect, and is also affected by the total energy stored in the system. For small $l$, the fringe effect becomes more serious as $l$ decreases, while the per-unit-length kinetic inductance remains more or less the same in that frequency regime, so the attractive pressure decreases. For large $l$, the fringe effect is negligible but the currents will decrease with an increase of $l$ as the Ohmic loss becomes more serious, so the pressure will also decrease. In either PEC or Au case, the resonant frequency red shifts with increasing $l$, as expected from Eq. (8). Fig. 2(c) shows the $d$-dependence of the optical pressure. As $d$ increases, the optical pressure drops quickly.

We present in Fig.3 the comparison between the numerical results and the analytical theory described in section II. In the PEC case, we use $\alpha$ in the analytical model as a fitting parameter to fit both the optical pressure and resonant wavelength at the same time and we find $\alpha=5.5$. Here $d_{eff}$ is just the gap distance between the plates ($d_{eff}=d=$10nm). In the plasmonic Au case, in addition to the fringe effect parameter $\alpha$, we introduce another fitting parameter $d_{eff}$. This is because the incident electromagnetic fields can penetrate the plates, which generate magnetic flux within the plates that should be included in the electromotive force calculation. And the fitting shows they take the values: $\alpha=38$ and $d_{eff}=40$nm. From the values of $\alpha$, it can be inferred that the fringe effect in plasmonic system is much stronger than that of the PEC system. This is also seen by examining our numerically computed field pattern.

So, we may conclude here that the electromagnetic forces are attractive at the micron scale in the metal sandwich system, and the main contribution to the attractive force is the kinetic inductance which is a consequence of field penetration into the Au metal plates.

## B. Electromagnetic forces at the millimeter scale corresponding to microwave frequencies

We now scale up the structure from the micron scale to the millimeter scale. The values of $d$ and $t$ are set to 0.1mm and 0.5 mm, respectively. For the system at the millimeter length scale, the corresponding resonant frequency is in the microwave regime. In this frequency range, we model Au with $\varepsilon_r = 1 + i\sigma_{Au}/(\varepsilon_0\omega)$ where the conductivity[48] $\sigma_{Au} = 4.098 \times 10^7 \text{S/m}$.

Figure 4 shows the numerically calculated electromagnetic pressures for both PEC plates and Au plates as a function of wavelength for various values of length $l$. As expected, the property of the PEC system is the same as that of the micron scale, except for a simple scaling of wavelength and $l$. This is because, under the assumption of PEC, the materials are non-dispersive, therefore the solutions to the Maxwell equations are scalable. While the Au system experiences an attractive electromagnetic force in the plasmonic regime (micron scale), the force becomes repulsive in this microwave regime. This is because Au behaves as a good conductor in the microwave regime, so its behavior resembles PEC. The major difference between PEC and Au is the Ohmic loss. Without dissipative loss, the electromagnetic pressure is a monotonic increasing function of $l$ as the quality factor increases with the size of the cavity (see the inset of Fig.4), but when the Ohmic loss is taken into account, the Au system possesses a maximum electromagnetic pressure of about $450\text{Pa}/(\text{mW}/\mu\text{m}^2)$ at a particular size parameter of $l=6$mm. We note that while the electromagnetic pressure is quite a bit smaller than that of the PEC, it is still a sizable one compared to the photon pressure of $6.67\text{Pa}/(\text{mW}/\mu\text{m}^2)$ exerted on a single PEC plate by a single plane wave (we note that the radiation pressure acting on other materials will be less than that of the PEC). And telling from Fig. 2(b) and Fig. 4, this value reaches about one fifth of the peak optical pressure in IR regime with the same intensity of the incident electromagnetic plane wave.

To analytically calculate the electromagnetic force in the microwave regime, we apply the same method as in the PEC case. One should notice that the total per-unit-length resistance now consists of two parts: $R = R_r + R_s$, where the per-unit-length surface resistance[47] takes the form $R_s = 2\sqrt{\mu_0 \omega / 2\sigma_{Au}}/w$ and $R_r$ is the per-unit-length radiation resistance. The first factor of 2 in $R_s$ is added because we have two inner surfaces that the induced currents flow on. We then follow the procedures in the PEC case to obtain the analytical value of electromagnetic force in the microwave cases.

Figure 5 compares the analytical results with the numerical results for the maximum electromagnetic pressure of the Au system in the microwave regime as a function of $l$. At each value of $l$, we do computation for a continuous range of wavelength to identify the resonant wavelength (see Fig. 4), which is shown as open (blue) circle in the inset of Fig.5, and the computed electromagnetic pressure at resonance for each value $l$ is displayed as open (blue) circle in the main panel of Fig. 5. The results obtained with the analytic model with a fitting parameter of $\alpha = 5.5$ are shown as solid (red) line in the inset for the resonant wavelength and in the main panel for the electromagnetic pressure and we note the excellent agreement for the resonant wavelengths and reasonably well agreement for the electromagnetic pressures. We note that the value of $\alpha$ here is the same as that of the PEC case in the micron regime, and the effective separation between the plates is just the gap distance ($d_{eff} = d = 0.1\text{mm}$) as field hardly penetrates the metal. The electromagnetic pressure has a maximum value for a certain value of $l$, and this is caused by fact that there are two kinds of resistance: the radiation resistance and the Ohmic resistance. For small $l$, radiation resistance dominates as the resonant frequency is high. The system basically behaves as a PEC system and the electromagnetic pressure increases with an increase of $l$ as we have shown in Fig. 4. For large $l$, the surface resistance dominates. Because it increases linearly with $l$ as in a serial circuit and the system becomes more lossy, the electromagnetic pressure decreases with $l$ in the large $l$ limit.

The rise and fall of the electromagnetic pressure as a function of $l$ can also be understood from the perspective of the quality factor $Q$, which may be obtained from the width of the peaks in Fig. 4. The $Q$ increases linearly with $l$ in the PEC case (where there is no Ohmic loss), while in the Au case (with Ohmic loss), there exists a maximum for $Q$. The total $Q$ of our system can be written as

$$\frac{1}{Q} = \frac{1}{Q_{rad}} + \frac{1}{Q_{abs}}, \tag{21}$$

where $Q_{rad}$ is due to the radiation and $Q_{abs}$ is due to the absorption in the metal (Ohmic loss). In the inset of Fig. 4 we show the computed $Q$ factors for both PEC (red dots) and Au (blue triangles) plates as a function of $l$ in the millimeter regime. We note for the PEC case the quality factor has a linear relationship with length $l$ as $Q_{rad} = al$, where $a \approx 15$ can be extracted from the fitting curve (solid black line). For a resonant cavity,

$$\frac{dU}{dt} = -\frac{\omega U}{Q} = -\omega U (\frac{1}{Q_{rad}} + \frac{1}{Q_{abs}}) = -(P_{rad} + P_{abs}), \tag{22}$$

where $P_{abs} = \omega U / Q_{abs}$ and $P_{rad} = \omega U / Q_{rad}$ are the time-averaged absorbed power and radiated power respectively. It is clear that for the same resonance mode $P_{abs} \propto U$. Combining with $\omega \propto l^{-1}$, we obtain $Q_{abs} = b/l$, where $b$ is a constant. Eq. (21) then becomes

$$Q = \frac{ab}{al + b/l}. \tag{23}$$

In Fig.4, we fit the numerical calculated quality factors for the Au case with Eq. (23) and obtain $b \approx 2000$. This reveals the dependence of $Q$ on length $l$ of the plates. The large value of $b$ indicates that magnitude of the electromagnetic force strongly depends on the Ohmic loss of the system in the microwave regime. If there were no Ohmic dissipation, the electromagnetic forces at resonance can reach giant values as the quality factor increases linearly with size in the PEC case. The radiation loss is only important when resonant frequency is relatively high, while the Ohmic loss can easily become the dominant loss mechanism in the low frequency region.

The electromagnetic force in the microwave regime can be written as

$$F = \frac{U\alpha}{2l} = \frac{P_c Q}{\omega} \frac{\alpha}{2l} \approx \frac{P_c Q \alpha}{2c\pi}, \tag{24}$$

where $P_c$ is the effective input power. Eq. (24) suggests that a good coupling of the input power into the cavity is desirable for achieving a large force, which is not surprising. The plane wave incidence we have considered is not necessarily an efficient way to couple energy into the cavity. One possible way to obtain good coupling is to use localized sources placed near the air gap. For example, one may employ two magnetic dipole sources placed at the ends of the plates instead of using a plane wave to excite the cavity. In Fig.6 (a) we show the numerically computed electromagnetic forces for both PEC and Au plates with two magnetic dipoles placed at the mouth of opening of the air gap. The powers of the dipole sources are normalized to $1 \text{W/m}$. The electromagnetic force per unit power is

$$\frac{F}{P_{dipole}} = \frac{F}{w/\omega\varepsilon_0} \approx \frac{\pi c \varepsilon_0 \alpha U}{2wl^2}, \qquad (25)$$

where $P_{dipole} = w/\omega\varepsilon_0$ is the effective radiation power of the magnetic dipole source (see Appendix). There is no simple way to calculate the $U$ in Eq. (25), but by fitting the numerically calculated forces in Fig. 6(a) we found that the peak value $F \propto l^3$ for the PEC case. We note that for plane wave excitation $F \propto l^2$ and so the force grows faster with $l$ if we employ dipole source excitation. Thus, combining Eq. (25), it indicates the total energy coupled into this system has a relationship $U \propto l^5$. Judging from the scaling laws, localized sources are more efficient than plane waves in coupling energy into the cavity and in inducing stronger forces. Since $U = \frac{L}{2}\int_0^l w^2 J_0^2 \sin^2\frac{\pi}{l}y dy = L J_0^2 w^2 l / 4$, we may conclude that $J_0 \propto l^2$ which means $H_0 \propto l^2$ ($H_0$ is the amplitude of the magnetic field at the center point of the system). This conclusion is confirmed by fitting the data of $H_0$ obtained through real simulations in Fig. 6(b).

For the Au system, the electromagnetic force does not increase monotonically with $l$ due to the Ohmic loss, as expected. Nevertheless, it is interesting to note that we can still obtain a measurable repulsive force (about $5.8\times10^{-3}\text{N/m}$ with two coherent magnetic dipole sources of 1W/m power). Consider a system with a dimension of $l\times w\times t = 14\text{mm}\times 30\text{mm}\times 0.5\text{mm}$ and $d = 0.1\text{mm}$, when each dipole source has a power of $0.1\text{W}$, we can obtain a force of about $5.8\times10^{-4}\text{N}$, which should be measurable.

The conclusion is that in the microwave regime, the force is repulsive and the force is primarily derived from the field leakage effect. It is also rather surprising that the microwave induced pressure in the metal sandwich is still significantly higher than the usual photon pressure, even though it is commonly believed that microwave photons cannot induce a significant electromagnetic force.

## C. Corrugated surfaces and rough surfaces

In previous sections, we treat the surfaces of the plates as if they were perfectly smooth. In reality, depending on the fabrication process, the metal surfaces inevitably contain some level of roughness. We shall see below that surface roughness and corrugation can potentially lead to attractive optical forces. This also suggests a way to engineer the optical force via surface roughness/corrugation. In the following, the metal plates are taken to have a length $l = 0.4 \mu m$, average thickness $t = 0.05 \mu m$ and average distance $d = 0.02 \mu m$.

Figure 7 shows the optical pressure at different levels of corrugation (or roughness) for both PEC and Au plates. In Fig. 7(a), we show the computed optical pressure for a sinusoidal corrugation with an amplitude of *h*, as shown in the left inset of Fig.7 (a). The surface profile on both sides of the plate has the form $\Delta(y) = h\sin(41\pi y/400)$, where $y \in [-200, 200]$ is the y-coordinate of the surface measured in nanometers, and $\Delta(y)$ is the deviation from the originally flat surface. In the main panel of Fig. 7(a), we show the numerically computed optical pressure for PEC plates as a function of wavelength for several values of *h*. As the amplitude of *h* increases, the resonant peak red shifts in the PEC case. More importantly, the optical pressure gradually changes sign from positive (repulsive) to negative (attractive) as the corrugation increases for the case of PEC. In the inset of Fig. 7(a), we show the optical pressures at resonance for Au plates for a few values of *h*. The sinusoidal surface profile also induces a red shift of the resonant frequencies (not shown) and the optical pressure becomes more negative so that the attractive forces increase as the amplitude of *h* increases for the case of Au.

We next consider randomly rough surfaces. Fig.7 (b) presents the same results as those in Fig. 7(a) except that we introduce random roughness on the surfaces instead of a sinusoidal corrugation. In the simulation, we apply the following surface profile generating function to produce a quasi-random rough surface:

$$\Delta(y) = \sum_{i=1}^{M} h_i \cos\left(\frac{n_i y}{100}\right), \tag{26}$$

where $h_i \in [-h, h]$ are random uniform deviates generated by a random number generator in the simulation, $h$ is the roughness amplitude, $M$ is taken to be 20, $n_i$ is another random number which is arbitrarily set to be $n_i \in [0, 200]$, and the y-coordinate of the surface is measured in nanometers. We sample $h_i$ and $n_i$ in such a way that the volume of the plate is approximately unchanged. The left inset of Fig. 7(b) shows a picture of the roughened plates, with the roughness drawn to scale. We find that the effect of random roughness is similar to that of the sinusoidal corrugation. Specifically, we see from the PEC results (dash lines in the main panel of Fig. 7(b)) that surface roughness induces red shifts of the resonant frequencies and the optical pressure at resonance shifts from positive gradually to negative as the surface roughness increases. The right inset of Fig. 7(b) shows the optical pressure at resonance computed for roughened Au plates, and the negative optical pressure increases in magnitude as the surface roughness increases. The trends are qualitatively the same for a sinusoidal surface profile (Fig. 7(a)) and a rough surface profile (Fig. 7(b)). This phenomenon may be understood by considering the strong localized electric field induced on the rough surfaces,[49,50,51] which increases the electric energy, and thus the attractive force. Alternatively, it may be understood from the view point of the "spoofed surface plasmons" created by the corrugation.[52] In this point of view, the corrugated PEC surfaces sustain "spoofed plasmons", and thus the results approach those of "plasmonic" plates progressively as the corrugation increases. So, the surface corrugation and roughness generally induce an attractive component to the optical pressure and we can induce or increase an attractive optical pressure simply by roughening or corrugating the surface. Although we have only shown results in the micron scale, the same conclusion will also hold for the PEC plates in other frequency regimes.

## IV. SUMMARY

The electromagnetic force/pressure acting on a pair of parallel metallic plates under electromagnetic illumination is considered at both the micron scale and millimeter scale.

The numerical computations are carried out using a boundary element method, which gives the solutions of the electromagnetic fields, and the Maxwell stress tensor approach, which gives the total force once the fields are known. We found that the metal plates would experience a sizable electromagnetic pressure that is two to three orders of magnitude stronger than the usual photon pressure if the metallic sandwich is at resonance with the incident electromagnetic wave. We found that the peak value of microwave induced pressure can reach about one fifth of the peak optical pressure in the IR regime with the same intensity of the incident electromagnetic plane wave. The system can be satisfactorily modeled as an open-end resonant transmission line after the fringe effect is taken into account.

At all length scales, the induced electric fields give attractive forces and the induced magnetic fields give repulsive forces and these two opposing effects tend to cancel each other. Strong forces can be obtained if the effect of one of the fields can be suppressed one way or another. In the high frequency (IR) regime, the magnetic field repulsion is suppressed by the shifting of the magnetic field energy into the kinetic energy of electrons, and that leads to an attractive force coming from the electric field. In the low frequency (microwave) regime, which is close to the PEC limit, the electric field leakage diminishes the attractive electric forces, leaving behind the repulsion due to magnetic field. The effect of surface corrugation and surface roughness is also investigated, and we find that corrugation/roughness generally induces attraction between the plates.

**ACKNOWLEDGMENTS**

This work is supported by Hong Kong Research Grants Council grant 600308. We would like to thank X. Q. Huang, Y. Lai and M. Y. Sun for the helpful discussions. S. B. Wang's studentship is supported by Nano Science and Technology Program, Hong Kong University of Science and Technology. H. Liu is financially supported by the National Natural Science Foundation of China (No.11074119, No.10874081). Computation resources are supported by the Shun Hing Education and Charity Fund and by the High Performance Cluster Computing Centre at Baptist University.

**APPENDIX**

**1 Symmetric mode and anti-symmetric mode in the parallel-plate system**

For the parallel-plate system, two kinds of coupling modes exist: the symmetric mode and the anti-symmetric mode. For the symmetric mode, the electric dipoles induced in the plates oscillate with the same phase. The induced currents have parallel directions and the localized electric charges on the same ends of the two plates have the same sign. In this case, the parallel currents exert attractive forces on the plates while the electric charges exert repulsive forces on the plates. The total force is then determined by the competition of the two kinds of forces, which is similar to the anti-symmetric mode we discussed in the main text. However, compared with the optical force in the anti-symmetric mode, optical force corresponding to the symmetric mode is much smaller in our cases. The symmetric mode gives a very small force on one plate, it can be more easily identified by showing the total optical pressure acting on the whole system (that is the sum of the pressures on both plates) defined as $P_{tot} = P_L + P_R$, where $P_L$ and $P_R$ are the optical pressures exerted on the left plate and the right plate, respectively. Fig. A1(a) shows the wavelength dependence of the total optical pressure. We note the symmetric mode has a fairly broad peak and there is higher order anti-symmetric mode as we have mentioned in the main text. Fig. A1(b) shows the optical pressure exerted on the right plate. We note the peak pressures corresponding to the anti-symmetric mode are so large compared with the symmetric mode that the symmetric mode can hardly be noticed.

**2 Telegraph equations for the transmission line model of our system under incident electromagnetic wave**

Under the illumination of electromagnetic wave with frequency $\omega$, the telegraph equations for the transmission line are (see Fig. A2)

$$\frac{dV(y)}{dy} + (R - i\omega L)I(y) = i\omega\mu_0 \int_{x_1}^{x_2} \hat{z} \cdot \mathbf{H}_{in}(x)dx, \tag{A1}$$

$$\frac{dI(y)}{dy} + (G - i\omega C)V(y) = i\omega(G + C)\int_{x_1}^{x_2} \hat{x} \cdot \mathbf{E}_{in}(x)dx. \tag{A2}$$

Here $\mathbf{H}_{in}(x)$ is the incident magnetic field and $\mathbf{E}_{in}(x)$ is the incident electric field. $R$ is the per-unit-length resistance of the metal plates, $G$ is the per-unit-length conductance of the medium sandwiched between the plates (zero in this case), $L$ and $C$ are the per-unit-length inductance and capacitance of the system, respectively and $d_{eff} = x_2 - x_1$ is the effective separation of the lines.

Detailed derivation of such equations can be found in the literature (see for example, Ref. 40). In the standard treatment of the problem, the metals are assumed to be good conductors. This can be extended to plasmonic systems by including the kinetic inductance, i.e., the total conductance becomes $L = L_m + L_e$, and an effective separation of the plates is defined to take care of field penetration into the metal plates. The final expressions of the telegraph equations remain formally the same.

## 3 Current distributions on the metal plates

Differentiating Eq. (A2) with respect to $y$, and then substituting it into (A1), we obtain the differential equation for $I(y)$ as

$$\frac{dI^2(y)}{dy^2} + A^2 I(y) = B, \qquad (A3)$$

where $A = \sqrt{\omega^2 CL + i\omega RC}$ and $B = -\mu_0 \omega^2 C \int_{x_1}^{x_2} \hat{z} \cdot \mathbf{H}_{in}(x) dx$. Solve (A3) with the boundary condition $I(0) = I(l) = 0$ (currents at the ends of the plates are zero) and we obtain

$$I(y) = \frac{B}{A^2}\left[1 - \cos(Ay) - \tan(Al/2)\sin(Ay)\right]. \qquad (A4)$$

Equation (A4) indicates the resonance condition is given by $Al = (2n+1)\pi$. At resonance it is expected $\tan(Al/2) \gg 1$, so we have $I(y) \approx -(B/A^2)\tan(Al/2)\sin(Ay)$. We then do a substitution with $A = \xi + i\zeta$, where $\xi = \text{Re}\sqrt{\omega^2 CL + i\omega RC}$ and $\zeta = \text{Im}\sqrt{\omega^2 CL + i\omega RC}$. And considering

$$\sin(Ay) = \frac{1}{2i}\left(e^{iAy} - e^{-iAy}\right) = \frac{1}{2i}\left(e^{-\zeta y}e^{i\xi y} - e^{\zeta y}e^{-i\xi y}\right), \tag{A5}$$

we rewrite the current distribution equation as

$$\begin{aligned}|I(y)|^2 &= I(y) \cdot I(y)^* \\ &= \frac{B^2}{|A|^4}|\tan(Al/2)|^2 \sin(Ay)\sin^*(Ay) \\ &= \frac{B^2}{4|A|^4}|\tan(Al/2)|^2 \left[e^{-2\zeta y} + e^{2\zeta y} - 2\cos(2\xi y)\right].\end{aligned} \tag{A6}$$

In the limit of $R \to 0, \zeta \to 0, A = \xi = \pi/l$. We have

$$|I(y)| = \frac{|B|}{2|A|^2}|\tan(Al/2)|\sqrt{2 - 2\cos(2\xi y)} = \frac{|B|}{|A|^2}|\tan(Al/2)|\sin\frac{\pi y}{l}, \tag{A7}$$

where $\sin(\pi y/l)$ represents the distribution effect of the transmission line. Eq. (A7) indicates that as long as the loss is small, we can always assume the current distributions on the plates are sinusoidal. In Fig. A3 we show the current distributions given by Eq. (A4) for three cases: $R=0$, $R=R_{plas}$ and $R=200R_{plas}$, where $R_{plas}$ is the actual per-unit-length resistance in our plasmonic system. Please note that for an ideal case of $R=0$ and $d/l \to 0$ the current amplitude given by Eq. (A7) should be infinitely large. However, in a real case with a non-zero $d/l$, the resonant current is finite. From the comparison in Fig. A3 we see the assumption of a sinusoidal current distribution in our plasmonic system is reasonable. And for the microwave regime this assumption should be even more accurate as the loss is smaller. We note the assumption gradually breaks down as the resistance of the system increases ($R=200R_{plas}$ case).

## 3 Derivation of the electromagnetic forces

With the fringe effect correction, the zero-th order resonant frequency is

$$\omega_0 = \frac{2\pi c}{\lambda} = \frac{\pi}{l\sqrt{LC(1+\alpha d/l)}}, \tag{A8}$$

where for the plasmonic system

$$L = L_m + L_e = \frac{\mu_0 d}{w} + \frac{1}{\varepsilon_0 \omega_p^2 \delta w}, C = \frac{\varepsilon_0 w}{d}. \tag{A9}$$

Combing Eq. (A8) and Eq. (A9), we deduce the expression for electromagnetic force in the plasmonic system at resonance. The resonance is dominated by one single mode with resonance frequency $\omega_0$ and $U$ is the electromagnetic energy stored in the system:

$$\begin{aligned} F &= -\frac{\partial U}{\partial d} \approx -\frac{\partial U}{\partial \omega_0}\frac{\partial \omega_0}{\partial d} \approx -\frac{U}{\omega_0}\frac{\partial \omega_0}{\partial d} \\ &= -\frac{U}{\omega_0}\frac{\partial \{\pi [LCl^2(1+\alpha d/l)]^{-\frac{1}{2}}\}}{\partial d} \\ &= \frac{U}{2}\frac{1}{(\mu_0 d/w + L_e)[(d/l)+\alpha(d/l)^2]}\frac{\mu_0 \alpha d^2 - L_e wl}{wl^2} \\ &\approx \frac{U}{2d}\frac{1}{(\mu_0 d/w + L_e)}\left(-L_e + \frac{\mu_0 \alpha d^2}{wl}\right) \\ &= \frac{U}{2d(L_m + L_e)}\left(-L_e + \frac{\mu_0 \alpha d^2}{wl}\right). \end{aligned} \tag{A10}$$

For the PEC, there is no contribution from the kinetic inductance ($L_e$) in Eq. (A10) and the corresponding electromagnetic force is

$$F \approx \frac{U}{2dL_m}\frac{\mu_0 \alpha d^2}{wl} = \frac{\alpha U}{2l}. \tag{A11}$$

**4 Time averaged radiation power of the magnetic dipole source**

In the BEM simulations, we define the magnetic dipole source through the magnetic field as

$$\mathbf{H}(\rho, \theta) = H_1^{(1)}(k\rho)\cos\theta \hat{z}, \tag{A12}$$

where $H_1^{(1)}(k\rho)$ is the first order of the first kind Hankel function, $k$ is the wavenumber, $\rho$ is the distance from the evaluation point to the source point and $\theta$ defines the angular dependence. Eq. (A12) defines the field value in the *x-y* plane and we assume no change in the third dimension. In the far-field limit Eq. (A12) becomes

$$\lim_{\rho \to \infty} \mathbf{H}(\rho,\theta) = \sqrt{\frac{2}{\pi k \rho}} e^{i(k\rho - \frac{3\pi}{4})} \cos\theta \hat{z}. \tag{A13}$$

The corresponding electric field is

$$\begin{aligned} \mathbf{E}(\rho,\theta) &= \frac{1}{-i\omega\varepsilon_0} \nabla \times \mathbf{H}(\rho,\theta) \\ &= \frac{-i\sin\theta}{\omega\varepsilon_0 \rho} \sqrt{\frac{2}{\pi k \rho}} e^{i(k\rho - \frac{3\pi}{4})} \hat{\rho} + \frac{k\cos\theta}{\omega\varepsilon_0} \sqrt{\frac{2}{\pi k \rho}} e^{i(k\rho - \frac{3\pi}{4})} \hat{\theta}. \end{aligned} \tag{A14}$$

The time averaged Poynting vector in the far-field can be written as

$$\mathbf{S} = \frac{1}{2} \mathrm{Re}(\mathbf{E} \times \mathbf{H}^*) = \frac{1}{\pi \omega \varepsilon_0} \frac{\cos^2\theta}{\rho} \hat{\rho}. \tag{A15}$$

Assume in the third dimension (along *z* axis) the source has the same length *w* with our parallel-plate system, we then integrate the above Poynting vector around a cylindrical surface surrounding the source to obtain the time averaged radiation power:

$$P = w \int_{\rho \to \infty} \mathbf{S} \cdot d\mathbf{\Omega} = \frac{w}{\omega \varepsilon_0}. \tag{A16}$$

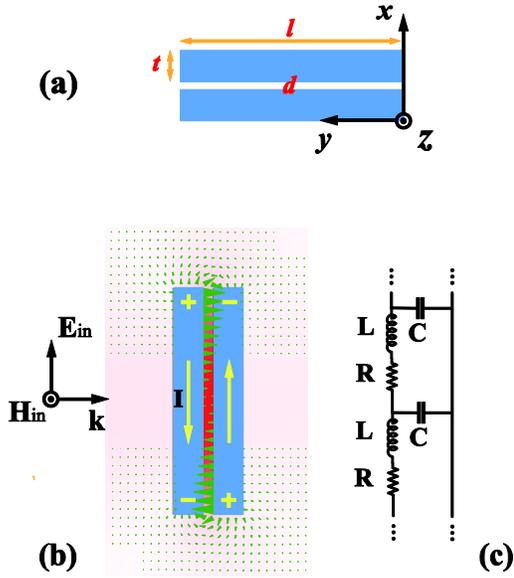

FIG.1. (Color online) A schematic of parallel-plate system. (a) Geometric parameters and the coordinate axis; (b) the direction of the incident plane wave and polarizations of $E_{in}$ and $H_{in}$ fields are indicated. The yellow arrows indicate pictorially the direction of the current $I$ flowing in the metal plates at one instant. For PEC, the currents would be surface currents. The numerically computed field distribution at the lowest order resonance for PEC plates ($l = 0.4\mu m, t = 0.05\mu m$ and $d = 0.02\mu m$) is shown to illustrate the field patterns at resonance. The magnetic field ($|H_z|$) (indicated by the red-colored pattern) is localized near the center region and the electric field (indicated by the green arrows) is the strongest in the vicinity of the open ends. (c) A transmission line model with per-unit-length $R$, $L$ and $C$ is used to analytically model our system.

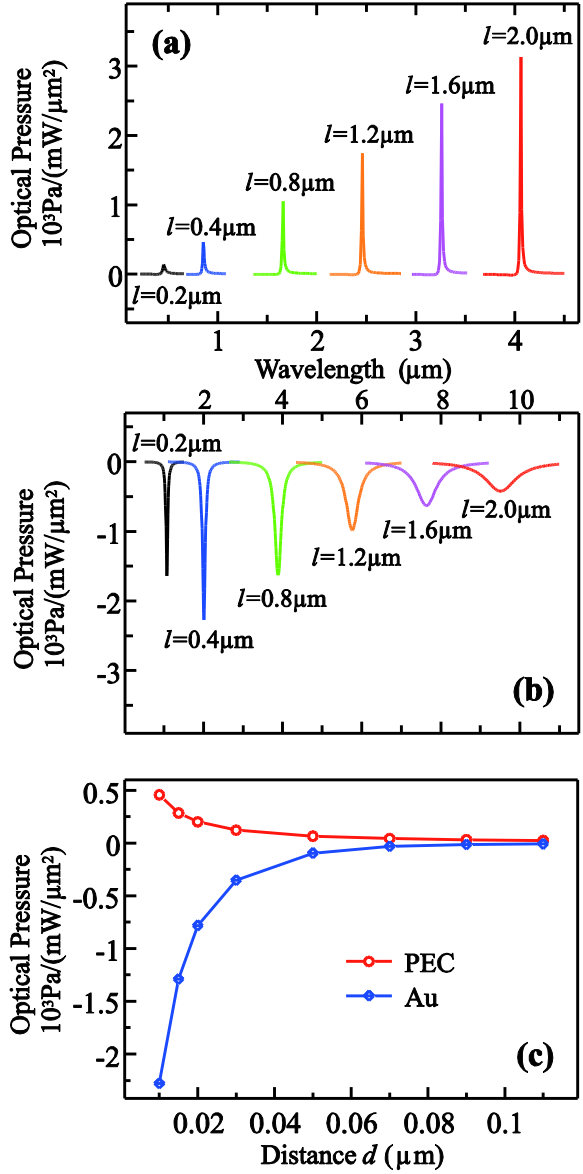

FIG.2. (Color online) Optical pressure in parallel-plate system at the micron scale as a function of wavelength for (a) PEC plates and (b) Au plates calculated using BEM and Maxwell stress tensor. In both (a) and (b) we have $t=0.05\mu m$ and $d=0.01\mu m$. (c) Optical pressure vs. distance $d$ between the plates with $t=0.05\mu m, l=0.4\mu m$.

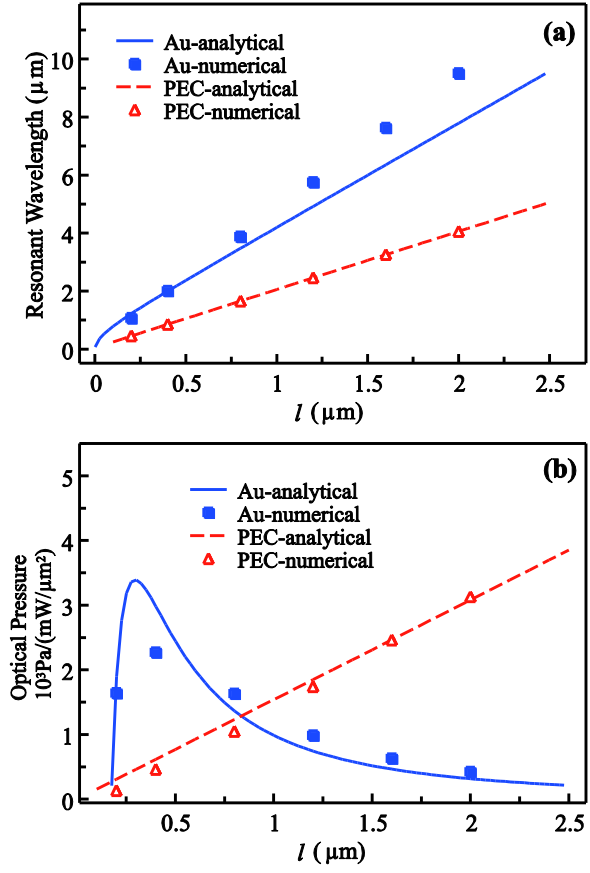

FIG.3. (Color online) Comparison of resonant wavelengths (panel (a)) and optical pressures (panel (b)) as a function of $l$ between the numerical results and the analytical theory. The squares (for Au) and triangles (for PEC) are calculated with BEM and the analytical results (solid lines for Au and dashed lines for PEC) are calculated with the transmission line theory. We note that in panel (b), we plot the absolute value of the negative pressure in the Au case for convenience. For the Au cases of (a) and (b) we have $\alpha = 38$, $d_{eff} = 40\text{nm}$, while for the PEC cases $\alpha = 5.5$, $d_{eff} = d = 10\text{nm}$.

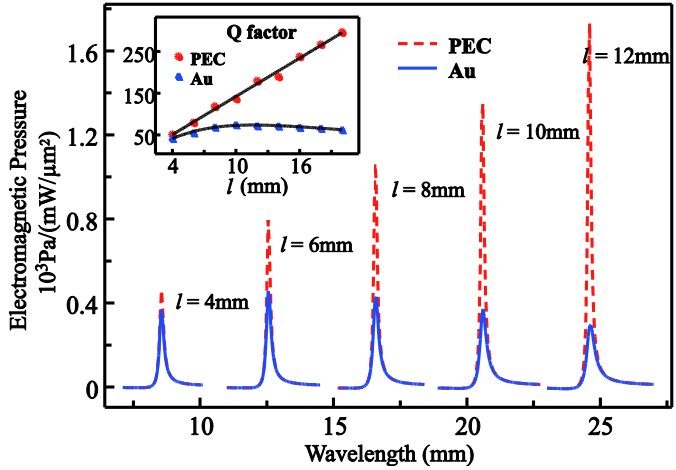

FIG.4. (Color online) Wavelength dependence of electromagnetic pressure for different values of $l$ in the millimeter regime, with the dotted red lines for PEC and solid blue lines for Au. The inset shows quality factors of the PEC system and the Au system (see text for details); the fitting curve for the quality factors for the PEC case is $y = 15x$ and for the Au case is $y = 3 \times 10^4 / (15x + 2 \times 10^3 / x)$.

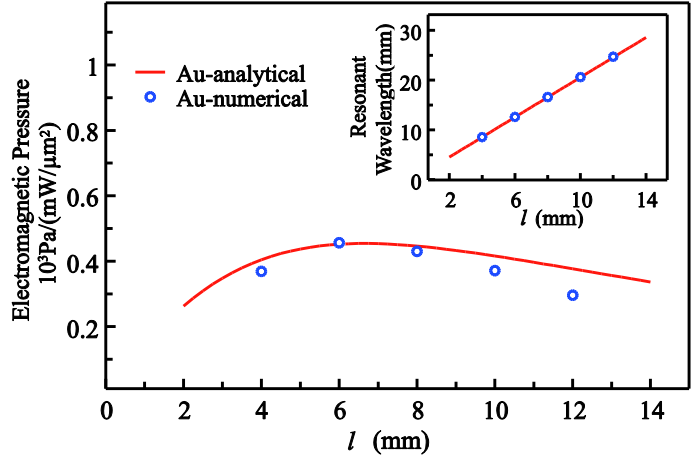

FIG.5. (Color online) Numerically calculated electromagnetic pressure (blue circles) as a function of $l$ compared with analytical theory (solid red lines) for Au system in the microwave regime. The inset shows comparison between numerically calculated resonant wavelength and the analytical theory. The fitting parameter is $\alpha = 5.5$.

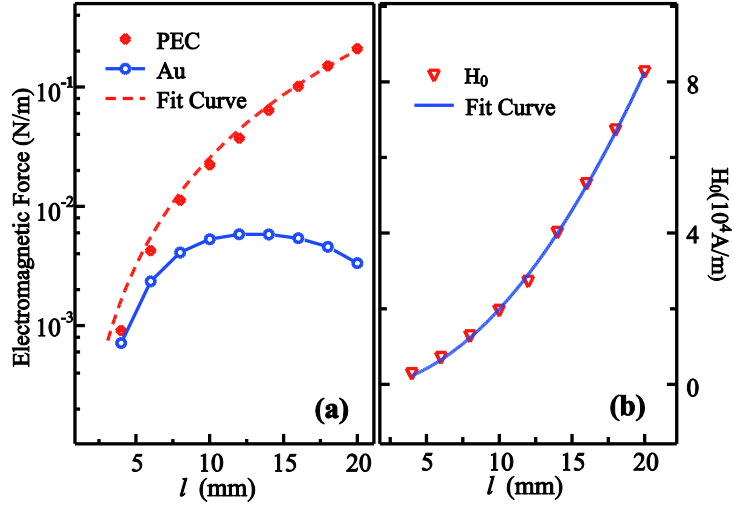

FIG.6. (Color online) (a) The electromagnetic force (in log-scale) induced by two coherent magnetic dipole sources of 1W/m; (b) Amplitude of the magnetic field in the center of the system. The fitting curve in (a) is $y = 2.55 \times 10^{-5} x^3$ and in (b) is $y = 209 x^2$. The simulation parameters are $d = 0.1\text{mm}, t = 0.5\text{mm}$.

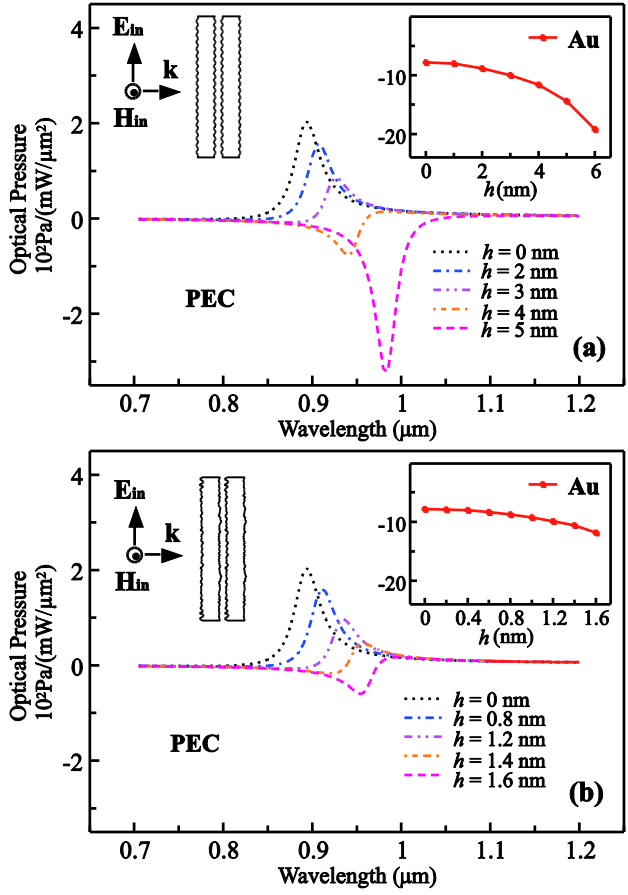

FIG.7. (Color online) In panels (a) and (b), we show the optical pressures for PEC plates as a function of wavelength for different values of $h$ in the main panels and the optical pressures at resonance for Au plates as a function of $h$ are shown in the right insets. In (a), $h$ is the amplitude of the sinusoidal corrugation and in (b) $h$ is the roughness amplitude defined in the text. The right insets have the same units as the main panels. A schematic of the corrugated/rough plates, drawn to scale, is shown on the left insets.

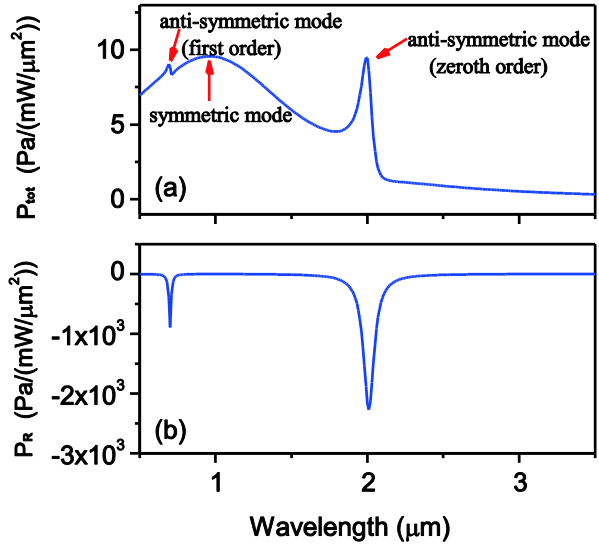

FIG. A1 (Color online) Symmetric mode and anti-symmetric mode in the parallel-plate Au system: (a) total optical pressure exerted on the whole system; (b) optical pressure exerted on the right plate. We set $l = 0.4\mu m, t = 0.05\mu m$ and $d = 0.01\mu m$.

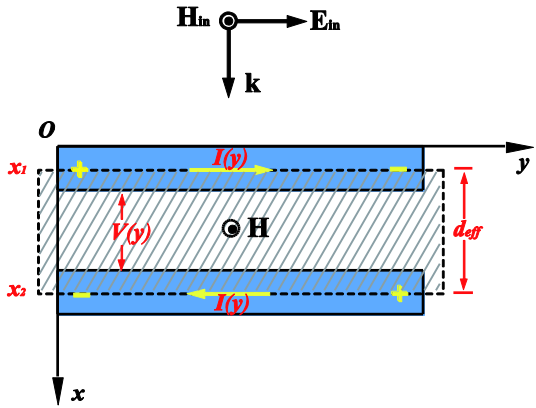

FIG. A2. (Color online) The transmission line model.

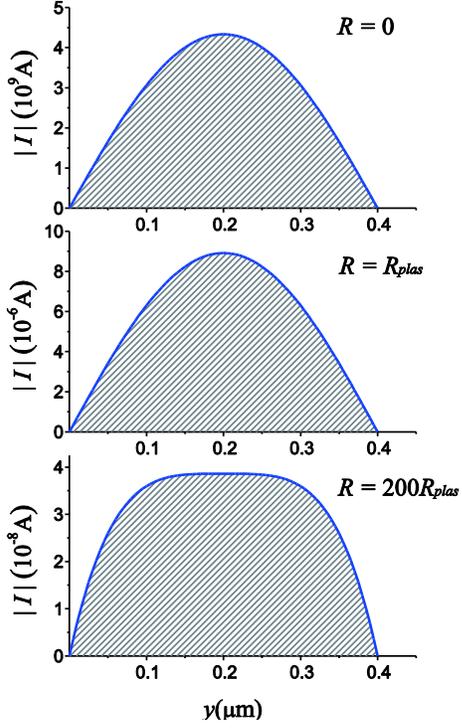

FIG. A3 (Color online) Current distributions on the metal plates for the cases of $R=0$, $R=R_{plas}$ and $R=200R_{plas}$ in the micron scale system under the illumination of an incident plane wave with intensity of $1mW/(\mu m)^2$. The relevant geometric parameters are: $l=0.4\mu m, t=0.05\mu m, d=0.01\mu m, w=100\mu m$.

*wangsb@ust.hk
†phchan@ust.hk